\documentclass[]{aa}

\usepackage{graphicx,natbib,multirow}

\newcommand{\msun}{\,M$_{\odot}$}
\newcommand{\rsun}{\,R$_{\odot}$}

\begin{document}

%\thesaurus{06(02.01.2;02.08.1;02.09.1;02.19.1;08.02.1;13.25.5)}

\title{A 2.4 - 12 $\mu$m spectrophotometric study with ISO of Cygnus\,X-3 in quiescence  
        \thanks{Based on observations with ISO,
        an ESA project with instruments funded by ESA Member States
        (especially the PI countries: France, Germany, the Netherlands
        and the United Kingdom) and with the participation of ISAS and NASA.}}
\author{Lydie Koch-Miramond\inst{1} 
\and P\'eter \'Abrah\'am\inst{2} \inst{,3} 
\and Ya\"el Fuchs\inst{1} \inst{,4} \and \\
Jean-Marc Bonnet-Bidaud\inst{1} 
\and Arnaud Claret\inst{1}}
\institute {DAPNIA/Service d'Astrophysique, CEA-Saclay, 91191 Gif-sur-Yvette Cedex, France
\and Konkoly Observatory, P.O. Box 67, 1525 Budapest, Hungary \and Max-Planck-Institut 
f\"ur Astronomie, K\"onigstuhl 17, D-69117 Heidelberg, Germany \and Universit\'e Paris VII, France
}
\titlerunning{Spectrophotometry of Cygnus X-3 in quiescence}
\offprints{Lydie Koch-Miramond (e-mail: lkoch@discovery.saclay.cea.fr)}
\date{Received 5 June 2002; Accepted 28 June 2002}
%\date{\today}

\abstract{
We present mid-infrared spectrophotometric results obtained with the 
ISO on the peculiar X-ray binary Cygnus X-3 in quiescence, at 
 orbital phases 0.83 to 1.04. The 2.4\,-\,12 $\mu$m continuum radiation  
observed with ISOPHOT-S can be explained by thermal free-free emission 
in an expanding wind with, above 6.5\,$\mu$m, a possible additional black-body 
component with temperature 
T\,$\sim$ 250\,K and radius R\,$\sim$\,5000{\rsun} at 10 kpc, likely due to 
thermal emission by circumstellar dust. 
The observed brightness and continuum spectrum closely 
match that of the Wolf-Rayet star WR 147, a WN8+B0.5 binary system, when 
rescaled at the same 10 kpc distance as Cygnus X-3. 
A rough mass loss estimate assuming a WN wind gives 
$\sim$ $1.2\times 10^{-4}$\,${\mathrm M}_\odot$.yr$^{-1}$. 
A line at $\sim$\,4.3 $\mu$m with a more than 4.3 $\sigma$ detection level, and 
with a dereddened flux of 126 mJy, is interpreted as the expected 
He I 3p-3s line at 4.295\,$\mu$m, a prominent line in the WR\,147 spectrum.  
These results are consistent with a Wolf-Rayet-like companion to the compact 
object in Cyg X-3 of WN8 type, a later type than suggested by earlier works.
}

\maketitle

\section{Introduction}
\indent
Cygnus\,X-3 has been known as a binary system since its discovery by
 \citet{gia67}, but there is still debate about the masses
of the two stars and the morphology of the system (for a review see \citealt{bon88}). 
The distance of
the object is 8-12.5\,kpc with an absorption on the line of sight 
A$_V$\, $\sim$\, 20 mag \citep{ker96}. The flux modulation 
at a period of 4.8 hours, first discovered in X-rays
\citep{par72}, then at near infrared wavelengths 
\citep{bec73}, and observed simultaneously at X-ray 
and near-IR wavelengths by \citet{mas86}, is believed to be the 
orbital period of the binary system. 
Following infrared spectroscopic measurements \citep{ker92},
where WR-like features have been detected in I and K band spectra, 
the nature of the mass-donating star is suggested to be a Wolf-Rayet-like star, 
but an unambiguous classification, similar 
to the other WR stars, is still lacking. \citet{mit96} and \citet{van98} 
pointed out that it is not possible to find
a model that meets all the observed properties of Cygnus\,X-3 where
the companion star is a normal Population I Wolf-Rayet star with a 
spherically symmetric stellar wind. In the evolution model originally 
proposed by \citet{heu73} a final period of the
order of 4.8 h may result from a system with initial masses 
M${_1^0}$\,=\,15\,{\msun}, M${_2^0}$\,=\,1\,{\msun}, P$^0$\,=\,5\,d, 
the final system being a neutron star
accreting at a limited rate of  $\sim$ 10$^{-7}$ \,{\msun}.yr$^{-1}$, 
 from the wind of a core He burning star of about 
3.8\,{\msun}. \citet{van98} proposed that the progenitor
of Cygnus X-3 is a 50\,{\msun}\,+\,10\,{\msun} system with P$^0$\,=\,6\,d; 
after spiral-in
of the black hole into the envelope of the companion, the hydrogen reach 
layers are removed, and a 2\,-\,2.5 {\msun} Wolf-Rayet like star remains
with P\,=\,0.2\,d. A system containing a black hole and an He core burning 
star is also favored  by \citet{erg98}.  
In addition, Cygnus\,X-3 undergoes giant radio bursts and 
there is evidence of jet-like structures moving away from Cygnus\,X-3 at 
0.3\,-\,0.9~c \citep{mio98,mio01,mar01}.\\
\indent
The main objective of the Infrared Space Observatory (ISO) spectrophotometric 
measurements in the 2.4\,-\,12 $\mu$m range was to constrain further
the nature of the companion star to the compact object: the expected 
strong He lines as well as 
the metallic lines in different ionization states are important clues,  
together with the spectral shape of the continuum in a wavelength range
as large as possible.     
An additional motivation for the imaging photometry with ISOCAM was to provide 
spatial resolution to a possible extended emission feature as a remnant of the 
expected high mass loss from the system.
The paper is laid out as follows. In Section 2 observational aspects 
are reviewed. Section 3 summarizes the results on the continuum and line 
emissions 
from Cygnus X-3 and four Wolf-Rayet stars of WN\,6, 7 and 8 types. 
Section 4 
reviews the constraints set by the present observations on the wind and 
on the nature of the companion to the compact object in Cygnux X-3. Finally, 
Section 5 summarizes the conclusions of this paper. 
\begin{table*}[!t] 
\centering 
\caption{The ISO observing modes, ISO identifications, wavelength ranges and observing times on
 April\,7, 1996, used on Cygnus\,X-3} 
\renewcommand{\arraystretch}{1.0} 
\setlength\tabcolsep{3.5pt} 
\begin{tabular}{llllll} 
\hline\noalign{\smallskip} 
Instrument & TDTNUM & Wavelength & Aperture  & \multicolumn{2}{l}{Total observing}  \\  
 &  & range ($\mu$m) & arcsec & time (s) & TU (start)   \\
\hline 
\noalign{\smallskip}
ISOCAM-LW10 & 14200701 & 8-15 & 1.5 & 1134 & 6:57:09 \\
\hline 
\noalign{\smallskip} 
ISOPHOT-SS &\multirow{2}{\jot}{14200803} & 2.4-4.9 & 24x24  & \multirow{2}{\jot}{4096} 
&\multirow{2}{\jot}{7:16:47} \\
ISOPHOT-SL & & 5.9-11.7 & 24x24  &   &\\
\hline 
\noalign{\smallskip}
ISOPHOT-P\,3.6 & \multirow{4}{\jot}{14200802} & 2.9-4.1 & 10  & \multirow{4}{\jot}{1100} 
&\multirow{4}{\jot}{8:27:33} \\
ISOPHOT-P\,10 & & 9-10.9 & 23  &  & \\
ISOPHOT-P\,25 &  & 20-29 & 52  &  &\\
ISOPHOT-P\,60 & & 48-73 & 99  &  &\\
\hline 
\end{tabular} 
\end{table*}
\section{Observations and data reduction\label{sobserv}}

We observed Cygnus\,X-3 with the Infrared Space Observatory 
(ISO, see \citealt{kes96}) on April\,7, 1996 corresponding to 
JD 2450180.8033 to 2450180.8519. 
The subsequent observing modes were: ISOCAM imaging photometry at
11.5\,$\mu$m  (LW10 filter, bandwith 8 to 15\,$\mu$m), 
ISOPHOT-S spectrophotometry in the range
2.4\,-\,12\,$\mu$m, for 4096\,s, covering the orbital phases
0.83 to 1.04 (according to the parabolic ephemeris of \citealt{kit92}); 
ISOPHOT multi-filter photometry at central wavelengths 
3.6, 10, 25 and 60\,$\mu$m. Observing modes 
and observation times are summarized in Table 1. Preliminary results 
were presented in \citet{koc01}. \\

\subsection{ISOPHOT-S data reduction}
A low resolution mid-infrared spectrum of Cygnus\,X-3 was obtained with
the ISOPHOT-S sub-instrument.
The spectrum covered the 2.4\,-\,4.9 and 5.9\,-\,11.7\,$\mu$m wavelength ranges 
simultaneously with 
a spectral resolution of about 0.04 and 0.1\,$\mu$m, respectively.
The observation was performed in the triangular chopped mode with two
background positions located at $\pm$120$''$, and with a dwelling time of
128\,s per chopper position. The field of view is 24{''}\,x\,24{''}.
The whole measurement consisted of 8 OFF1--ON--OFF2--ON cycles and lasted
4096\,s.

The ISOPHOT-S data were reduced in three steps. We first used the Phot 
Interactive Analysis (PIA\footnote
{PIA is a joint development by the ESA Astrophysics Division and the ISOPHOT
consortium led by the Max--Planck--Institut f\"ur Astronomie, Heidelberg.}, 
\citealt{gab97}) software (Version 8.2) to filter out cosmic glitches in the
 raw data and to determine signals by performing linear fits to the integration ramps.
After a second deglitching step, performed on the signals, a dark current value
appropriate to the satellite orbital position of the individual signal was subtracted.
Finally we averaged all non-discarded (typically 3) signals in order
to derive a signal per chopper step.
Due to detector transient effects, at the beginning of the observation
the derived signals were systematically lower than those in the consolidated
part of the measurement.
We then discarded the first $\sim$800\,sec (3 OFF-ON transitions), and 
determined an average [ON--OFF] signal for the whole measurement by applying
a 1-dimensional Fast Fourier Transformation algorithm (for the application
of FFT methods for ISOPHOT data reduction see \citealt{haa00}). The [ON--OFF] 
difference signals were finally calibrated by applying a signal-dependent 
spectral response function dedicated to chopped ISOPHOT-S observations
\citep{aco01}, also implemented in PIA.

In order to verify our data reduction scheme (which is not completely
standard due to the application of the FFT algorithm) and
to estimate the level of calibration uncertainties, we reduced HD\,184400,
an ISOPHOT standard star observed in a similar way as Cygnus\,X-3.
The results were very consistent with the model prediction of the star, 
and we estimate that the systematic uncertainty of our
calibration is less than 10$\%$.

\subsection{ISOPHOT spectral energy distribution}
\begin{figure*}[!htp] 
\centering
%\vspace*{0.9cm}
\includegraphics[width=13.cm,angle=0]{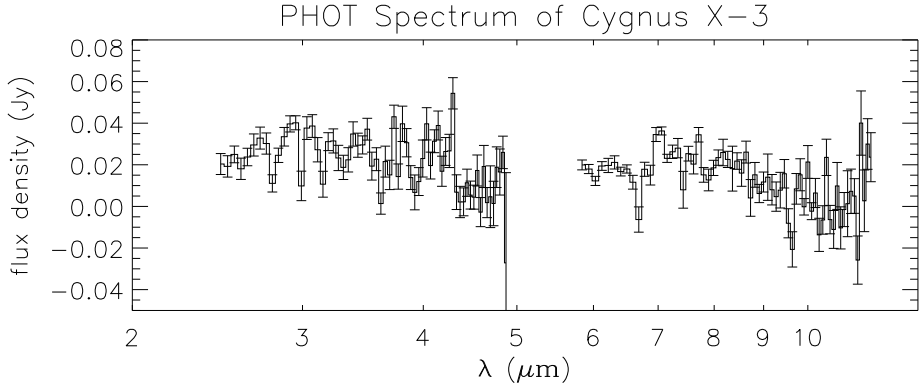}
%\vspace*{0.5cm}
\caption{Observed spectrum of Cygnus\,X-3 in the 2.4\,-\,12\,$\mu$m range, obtained 
with the FFT method.}  
\label{Fig1} 
\end{figure*}

The observed spectral energy distribution is shown on Fig.~1. The observed 
(not dereddened) continuum flux in the range 2.4\,-\,7\,$\mu$m is 20\,$\pm10$\,mJy 
in good agreement 
with that observed by \citet{ogl01} with ISOCAM on the same day (the dereddened 
fluxes are shown on Fig. 2) ; 
the observed flux decreases to about 10\,$\pm8$\,mJy around 9\,$\mu$m.

An unresolved line is observed at about 4.3 $\mu$m peaking at 57\,$\pm$\,10 mJy. 
The linewidth is 0.04\,$\mu$m, consistent with the instrumental
response and corresponding to $\sim$\,2500 km.s$^{-1}$. Note that the measured 
line flux might be underestimated because the ISOPHOT-S pixels are 
separated by small gaps, and a narrow line might falls into a gap.  

\subsection{ISOPHOT-P data analysis and results}
The data reduction in the multi-filter mode was performed using the Phot Interactive 
Analysis \citep{gab97} 
software. After corrections for non-linearities of the integration ramps, the signal 
was transformed to a standard reset interval. Then an orbital dependent dark current 
was subtracted and cosmic ray hits were removed. In case the signal did not fully 
stabilize during the measurement time due to the detector transients, only the last 
part of the data stream was used. The derived flux densities were corrected for the 
finite size of the aperture by using the standard correction values as stated in the 
ISOPHOT Observer's Manual \citep{kla94}.
The flux detected at 3.6\,$\mu$m 
(bandwidth 1\,$\mu$m) in the $10''$ diameter aperture
 is 8.1\,$\pm$\,3.3 mJy at a confidence level of 2.4\,$\sigma$.
 No detection above the galactic noise was obtained at 10, 25 and
60\,$\mu$m with $23''$, $52''$ and $99''$ diameter aperture, respectively.

\subsection{ISOCAM data reduction and results}
 The LW10 filter centered at 
11.5\,$\mu$m was used with the highest spatial resolution of $1.5''\times1.5''$ per pixel. 
The ISOCAM data were reduced with the Cam Interactive Analysis software (CIA\footnote
{CIA is a joint 
development by the ESA Astrophysics Division and the ISOCAM 
consortium. The ISOCAM consortium was led by the ISOCAM PI, C. 
Cesarsky, Direction des Sciences de la Matière, C.E.A, France.}) 
version 3.0, following the
standard processing outlined in \citet{sta99}. 
First a dark correction was applied, then a de-glitching to remove 
cosmic ray hits, followed by a transient correction to take into account 
memory effects, using the inversion algorithm of \citet{abe96}, and a flat-field 
correction. Then individual images were combined into the final raster map, 
whose pixel values were converted into milli-Jansky flux densities. No colour correction 
was applied. A point source is clearly visible at the Cygnus X-3 position 
on the ISOCAM map at 11.5\,$\mu$m. The measured flux is 7.0\,$\pm$\,2.0\,mJy
 above a uniform  background at a level of about 
1.2\,mJy, in good agreement with our ISOPHOT result. This flux is lower 
than the 15.2\,$\pm$\,1.6\,mJy (at 11.5\, $\mu$m) measured 
by \citet{ogl01}, on the same day, using ISOCAM/LW10 with a $6''\times 6''$ aperture. 
%No evidence for the presence of an extended source has been observed in our data
%when fitting the point spread function to the source profile. 

The high resolution configuration of the ISOCAM camera has been used to
constrain the spatial extension of the infrared source.
The measured FWHM for the source is 3.90\,$\pm$\,0.45\,arcsec (mean value of the 
four individual images composing the final raster map).
This can be compared to the ISOCAM catalogued point spread functions
at these energy and configuration which show a FWHM mean
value of 3.44\,$\pm$\,0.45 arcsec,
including the effects of the satellite jitter and of the pixel sampling.
The slightly larger value for Cygnus X-3, though only marginally
significant, might therefore indicate an extended source. The deconvolved 
extension would be 1.84 $\pm$ 0.64 arcsec which
at a distance of 10 kpc corresponds to a linear extension of $\sim$ $2.7\times10^{17}$\,cm.
The extended infrared source may be the result of the heating of the
surrounding medium by the radio jets whose existence have been now 
clearly demonstrated both at arcsec  \citep{mar00, mar01} and sub-arcsec 
\citep{mio01} scales, but it clearly deserves confirmation.

\section{Results and discussion\label{sresult}}
\subsection{Continuum spectral energy distribution\,:\,model fitting and comparison 
with four Wolf-Rayet stars}

\begin{figure*}[!htp] 
\vspace*{0.5cm}
\centerline{\hspace*{1.0cm}
\includegraphics[height=5.5cm]{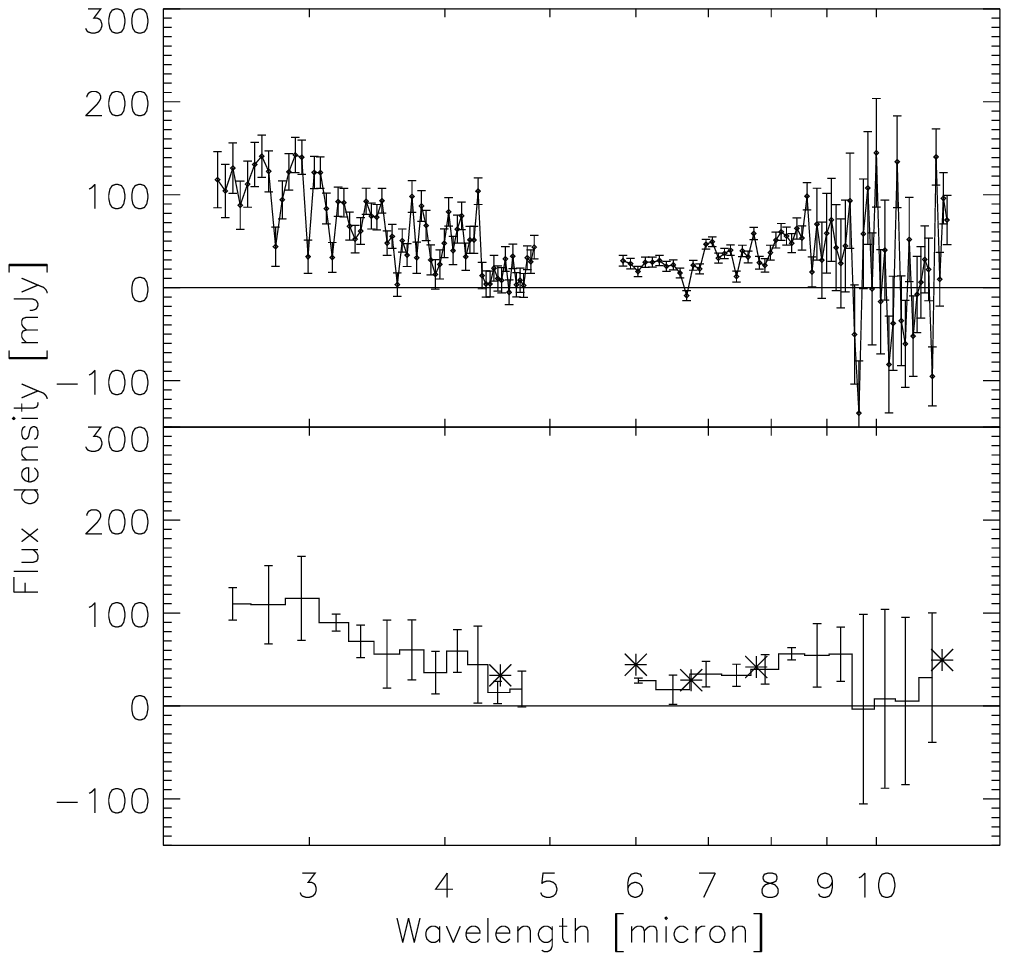}
\hspace*{2.cm}
\includegraphics[height=5.5cm]{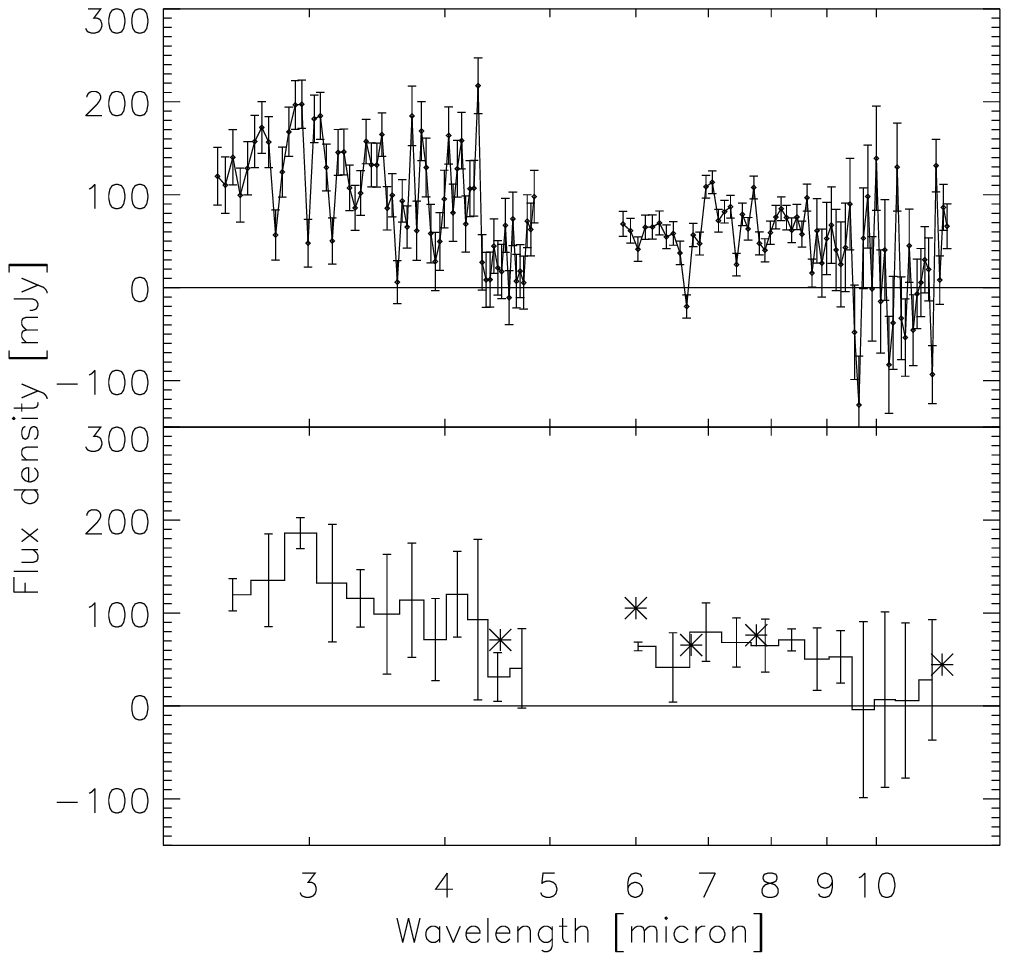}
}
\vspace*{0.8cm}
\caption{Dereddened spectrum of Cygnus X-3 using either \citet{dra89} law (left) or 
\citet{lut96} law (right). The asterisks (*) in 
the lower panel represent the \citet{ogl01} ISOCAM results, after dereddening 
by \citet{dra89} (left) and Lutz (right).}
\label{Fig2} 
\end{figure*}
\begin{figure*}[!htp]
\centerline{
\includegraphics[width=8cm]{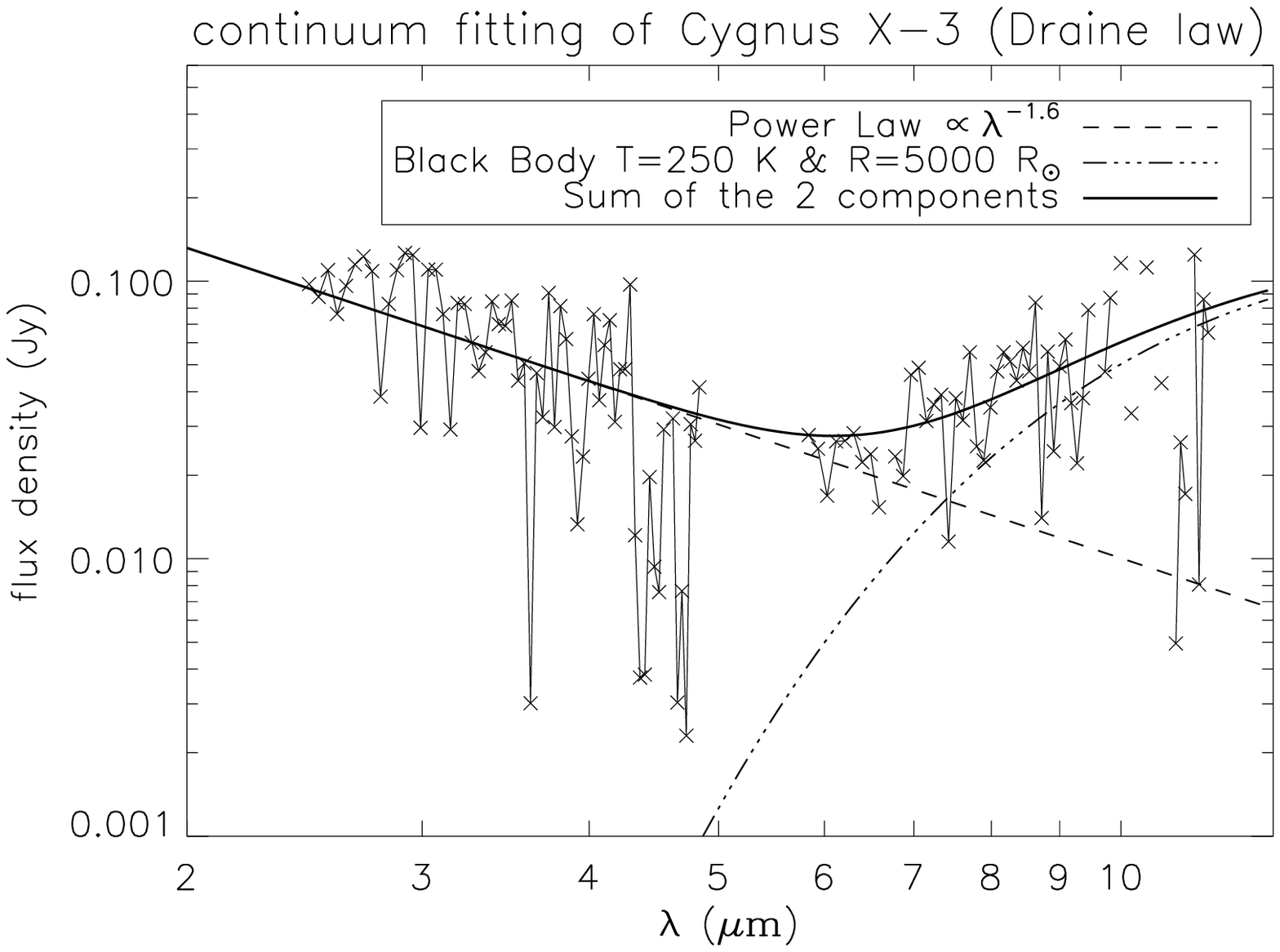}
\hspace*{0.5cm}
\includegraphics[width=8cm]{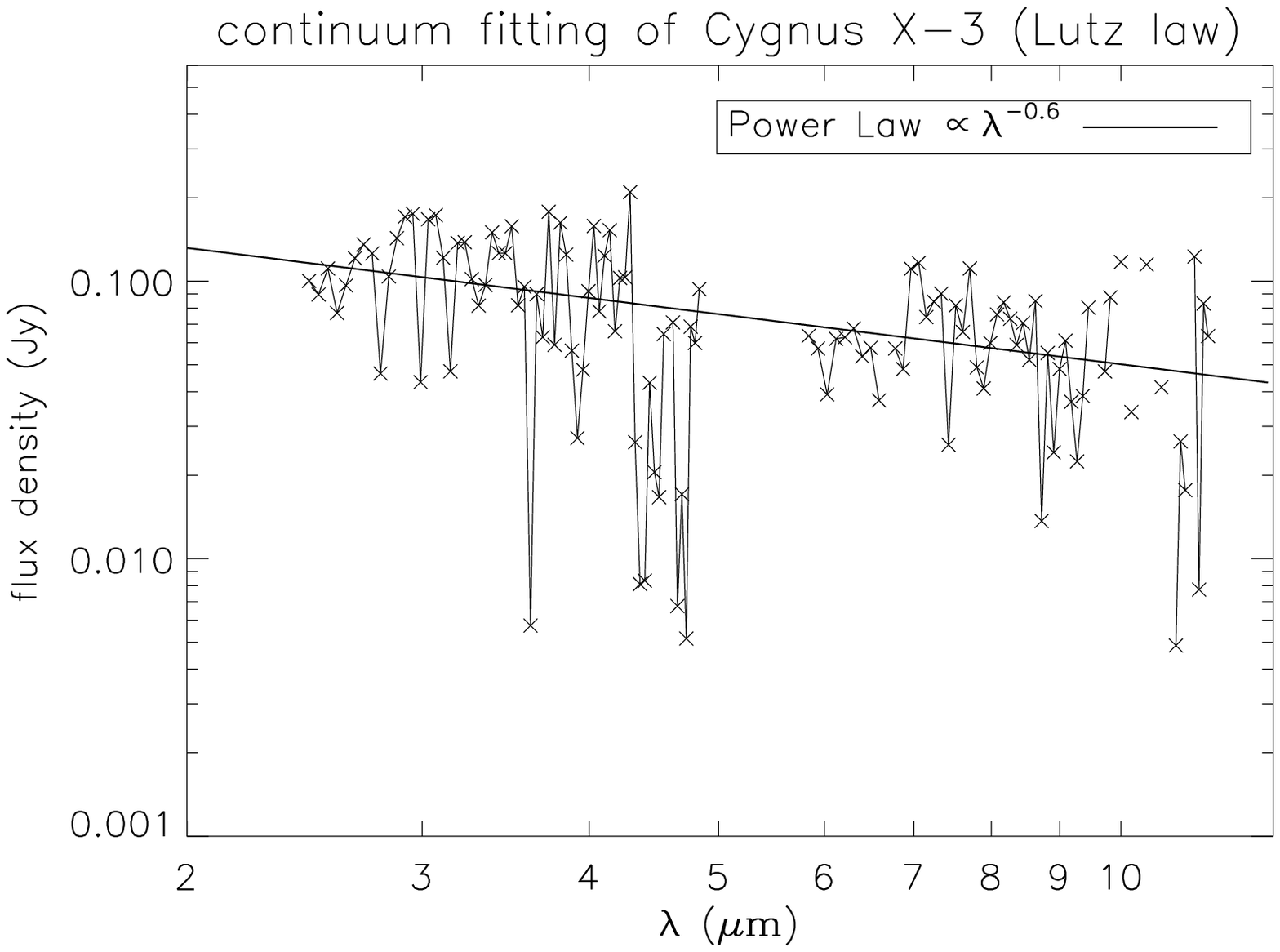}
}
\caption {Best fitting of the two dereddened spectra of Cygnus X-3 shown on Fig. 2.}  
\label{Fig3} 
\end{figure*}
	The dereddening of the Cygnus X-3 spectrum is made using
	either \citet{lut96} or \citet{dra89} laws, and using an
	absorption value of A$_V$\,=\, 20 mag \citep{ker96}; the two
	dereddened spectra are shown in Fig.~2. They have clearly
	different shapes, but since the molecular composition of the
	absorbing material on the line of sight to Cygnus\,X-3 is
	unknown, we cannot choose between these two laws.\\ 
   \indent The spectral fitting of the dereddened spectra is shown in
   Fig. 3 and the results are given in Table 2.  With the
   \citet{lut96} law the best fit is obtained with a unique power law
   $S_\nu$\,$\propto$\,$\lambda$$^{-0.6\pm}$\,$^{0.3}_{0.4}$ with a
   reduced $\chi^2$\,=\,4.3 in the 2.4-12\,$\mu$m range, in good
   agreement with the Ogley et al (2001) result.  With the
   \citet{dra89} law the best fit is obtained with the sum of two
   components: a power law with slope $\lambda$$^{-1.6\pm}$\,$^{0.2}$
   and a black body at T\,=\,250 K with a radius of 5000\,{\rsun} at a
   distance of 10\,kpc, (reduced $\chi^2$\,=\,4.2 between 2.4 and
   12\,$\mu$m), a hint of the presence of circumstellar dust.  The
   power law part of the continuum spectrum
   ($S_\nu$\,$\propto$\,$\lambda$$^{-\alpha}$) can be explained by
   free-free emission of an expanding wind in the intermediate case
   between optically thick ($\alpha$\,=\,2) and optically thin
   ($\alpha$\,$\sim$\,0) regimes \citep{wri75}.
 
\begin{table}[!h]
\centering 
\caption{Comparison of the infrared continuum spectra of Cygnus\,X-3 and four 
WR stars: power law slopes $\alpha$ such as S$_\nu$ $\propto$ $\lambda$$^{-\alpha}$ 
and 4.7 $\mu$m flux densities rescaled at 10 kpc} 
\renewcommand{\arraystretch}{1.0} 
\setlength\tabcolsep{5.5pt} 
\begin{minipage}{\hsize}
\begin{tabular}{@{}llll@{}} 
\hline
\noalign{\smallskip}
Object & PL slope\footnote{dereddening with \citet{dra89} law } 
    & PL slope\footnote{dereddening with \citet{lut96} law; fit between \mbox{2.4\,-\,12\,$\mu$m}}
 & Flux at 4.7$\mu$m (Jy)  \\
\noalign{\smallskip}
\hline
\noalign{\smallskip}
Cygnus\,X-3 & 1.6 $\pm$ 0.2 \footnote{fit between 2.4\,-\,6.5 $\mu$m} &  0.6 $\pm$ $^{0.3}_{0.4}$
& 0.079 $\pm$ 0.011 \\
WR\,78 & 1.4 $\pm$ 0.2 \footnote{fit between 2.4\,-\,12 $\mu$m} & 1.4 $\pm$ 0.2 
& 0.114 $\pm$ 0.008 \\ 
WR\,134  & 0.2 $\pm$ 0.2 $^d$  &  0.2 $\pm$ 0.2 & 0.081 $\pm$ 0.006 \\ 
WR\,136  & 1.0 $\pm$ 0.2 \footnote{fit between 2.4\,-\,8 $\mu$m}&  1.0 $\pm$ 0.2 
& 0.076 $\pm$ 0.006\\
WR\,147 & 1.6 $\pm$ 0.1 $^c$ &  1.0 $\pm$ 0.1 & 0.085 $\pm$ 0.005 \\
\hline 
\end{tabular}   
\end{minipage}
\end{table}

\begin{table*}[!htp] 
\centering 
\caption{The four Wolf-Rayet stars observed with ISO/SWS } 
\renewcommand{\arraystretch}{1.0}  
\begin{minipage}{\hsize}
\begin{tabular}{@{}lllllll@{}} 
\hline
\noalign{\smallskip}
Star  &Type & Binarity & Distance & A$_V$\footnote{from \citet{huc01} except for WR 147}
 & Reference  & TDTNUM\\
\noalign{\smallskip}
\hline
\noalign{\smallskip}
WR\,78 & WN7h$^a$ WNL & No & 2.0\,kpc & 1.48-1.87 & \citet{cro95II}  & 45800705 \\ 
WR\,134  & WN6 & possible & $\sim$ 2.1\,kpc & 1.22-1.99 & \citet{more99}  & 17601108 \\ 
WR\,136  & WN6b(h) WNE-s\footnote{from \citet{cro95IV}} & possible 
& 1.8\,kpc & 1.35-2.25 & \citet{ste99}  & 38102211 \\
WR\,147 & WN8(h) WNL & B0.5V at 0.554$''$ & 630 $\pm70$\,pc & 11.2 &\citet{morr99,morr00} & 33800415 \\
\hline 
\end{tabular} 
\end{minipage}
\end{table*} 
\begin{figure*}[!htp]
\vspace*{1.cm} 
\centering
\includegraphics[width=10cm]{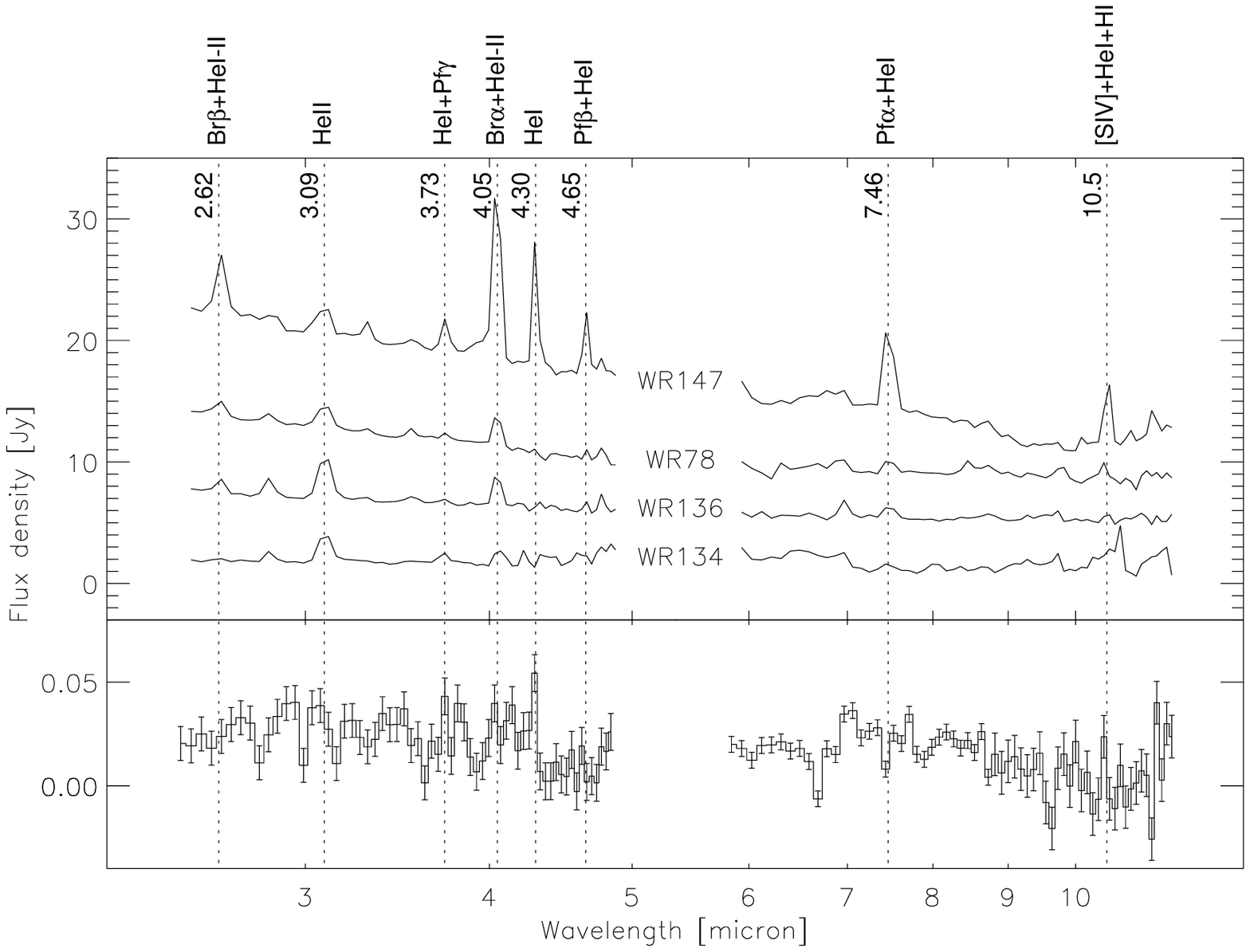}
\vspace*{1.0cm} 
\caption{Observed spectra of Cygnus\,X-3 and four Wolf-Rayet stars (not dereddened). An 
arbitrary vertical offset has been added to the WR spectra for clarity. The
 identification of the emission lines is from \citet{morr00}. }
\label{Fig4} 
\end{figure*}

	Using the ISO archive data we have analysed the SWS
	spectra of four Wolf-Rayet stars: WR\,147 (WN8+B0.5), WR\,136
	(WN6b), WR\,134 (WN6) and WR\,78 (WN7) whose main
	characteristics are given in Table\,3. We compare them to the
	Cygnus X-3 spectrum, after smoothing the SWS spectra to the
	resolution of the ISOPHOT-S instrument (using an IDL routine
	of B. Schulz dowloaded from the Home Page of the ISO Data
	Centre at Vilspa). The observed WR spectra are shown on Fig.
	4 on top of the observed Cygnus\,X-3 spectrum; the
	identification of the emission lines is from \citet{morr00}.

   The dereddened spectra of the Wolf-Rayet stars, using
   either the \citet{dra89} law or the \citet{lut96} law with the
   A$_V$ shown in Table\,3, have been fitted with power law slopes
   given in Table 2. Wolf-Rayet stars emit free-free continuum
   radiation from their extended ionized stellar wind envelopes and
   the different slopes reflect different conditions in the wind
   \citep{wil97}. It is noticeable that the mean continuum flux
   density of \mbox{Cygnus\,X-3} is the same (within a factor 1.5 at
   4.7\,$\mu$m as seen in Table~2) as that of the four WR stars when
   their flux density is rescaled to a Cygnus\,X-3 distance of
   10\,kpc.

	The comparison between the Cygnus\,X-3 spectrum and
	that of the Wolf-Rayet WR\,147 at 10\,kpc is shown in Fig. 5
	after dereddening with the \citet{dra89} law (left) and with
	the \citet{lut96} law (right).  The WR\,147 spectrum appears
	as the closest WR one to the Cygnus\,X-3 spectrum, with almost
	the same mean flux density at 10\,kpc and the same power law
	slope (whithin the statistical errors).
 
\begin{figure*}[!htp]
\vspace*{0.5cm} 
\centerline{
\includegraphics[width=8cm]{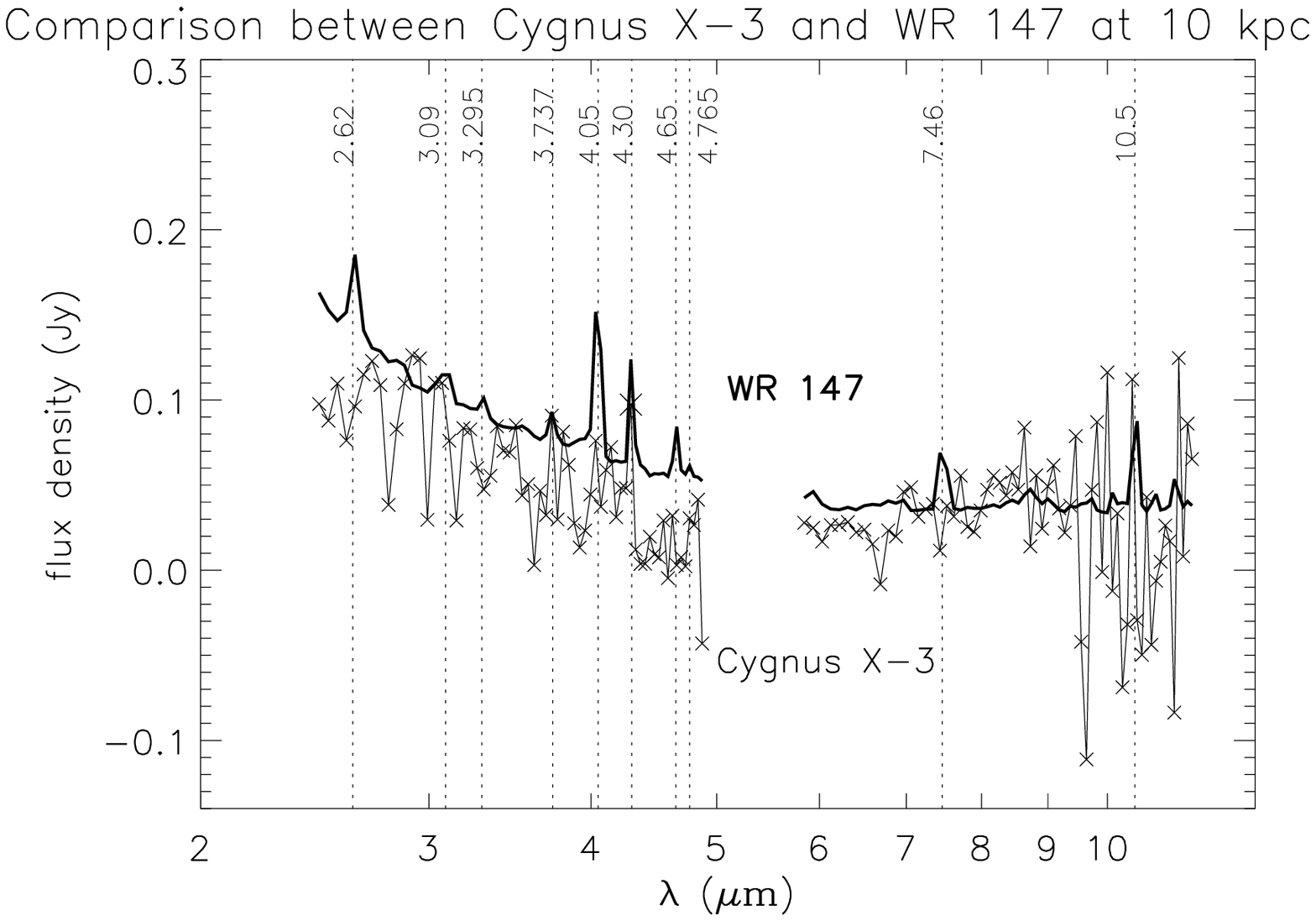}
\hspace*{0.5cm}
\includegraphics[width=8cm]{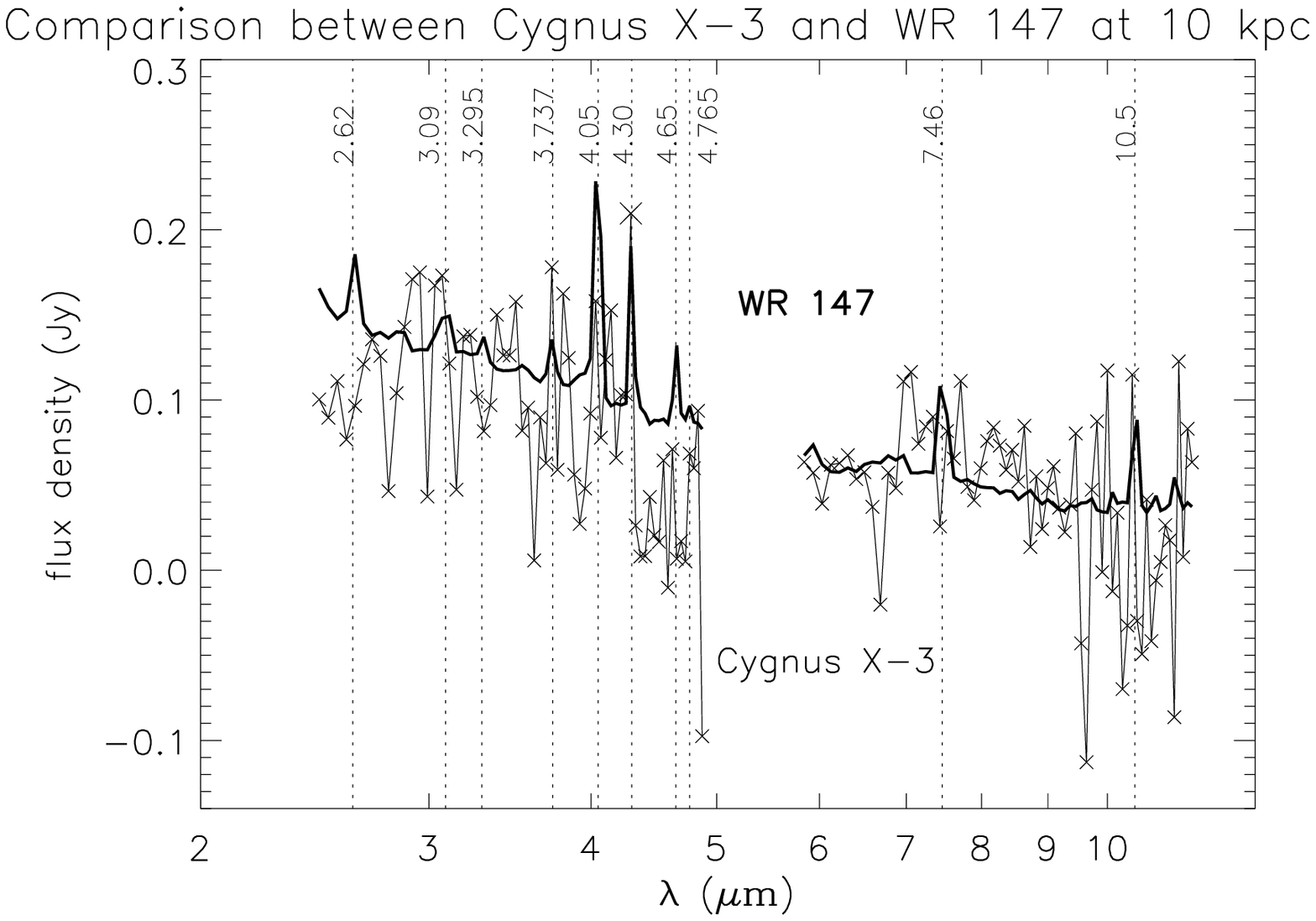}
}
\vspace*{0.1cm}
\caption{Comparison between the spectral energy distribution of Cygnus\,X-3 and
the one of WR\,147 when rescaled at 10 kpc : both dereddened with 
\citet{dra89} law (left) and the \citet{lut96} law (right).}
\label{Fig5} 
\end{figure*}

We note that WR\,147 is known as a colliding-wind binary that has been 
spatially resolved \citep{wil97,ski99}, with a separation on the sky large 
enough for the wind-wind collision 
zone between the stars to be resolved at near-infrared and radio \citep{wil97}, 
and X-ray energies \citep{pit02}. The spectral energy distribution of 
WR\,147 in the 0.5\,$\mu$m to 2\,mm wavelength range 
(including all components) shown by \citet{wil97} is dominated by the free-free 
emission from the stellar wind of the WN8 star; in the 2 to 10\,$\mu$m
range these authors find $\alpha$\,=\,1.0, in good agreement with our 
ISOPHOT-S measurement (when dereddened with the \citealt{lut96} law); and in 
the mid-infrared to radio range they find $\alpha$\,=\,0.66.

\subsection{Emission lines; comparison with four Wolf-Rayet stars}

   The measured 4.3\,$\mu$m line flux above the continuum in the
   Cygnus\,X-3 spectrum is 58\,$\pm$\,11\,mJy (dereddening with the
   \citealt{dra89} law), and 126\,$\pm$\,25\,mJy (dereddening with the
   \citealt{lut96} law), using respectively $\alpha$\,=\,1.6 and
   $\alpha$\,=\,0.6, the best fitted continuum slopes as given in
   Table~3, both detections being at more than 4.3\,$\sigma$.  This
   line is interpreted as the HeI (3p-3s) line at 4.295\,$\mu$m, a
   prominent line in the WR\,147 (WN8+B0.5) spectrum as seen in
   \citet{morr00} and in Fig.\,4.  Again WR 147 appears as the closest
   WR to Cygnus X-3 as being the only WR in our sample with a HeI
   emission line at 4.3\,$\mu$m, the only line clearly seen in our
   Cygnus\,X-3 data.  The other expected He lines at 2.62, 3.73, 4.05,
   7.46 and 10.5\,$\mu$m are not detected, probably due to the
   faintness of the object, at the limits of the instrument's
   sensitivity.  We note (Fig. 4) that the second highest peak in the
   SS-part of the Cyg\,X-3 spectrum is at 3.73\,$\mu$m, and there are
   also local maxima at 4.05 and 10.5\,$\mu$m. These expected lines
   are all blended with H lines and the absence of H observed by
   \citet{ker96} in the I and K band spectra of Cygnus\,X-3 could
   explain the weakness or absence of these lines in our data. 

We note that the Br$\alpha$+HeI-II line at 4.05\,$\mu$m is not
detected in Cygnus\,X-3 in quiescence, but is present in the four
Wolf-Rayet stars. Strong HeI and HeII lines have been previously
observed in the K-range in \mbox{Cygnus\,X-3} during quiescence
\citep{ker92,fen96,fen99}.  These lines have been interpreted
\citep{ker96,che94}, as emission from the wind of a massive companion
star to the compact object, and \citet{fen99} suggest that the best
candidate is probably an early WN Wolf-Rayet star.  We note that the
close match we have found between the mid-infrared luminosity and the
spectral energy distribution, the HeI emission line in Cygnus\,X-3 in
quiescence, and that of the WR\,147, is consistent with a
Wolf-Rayet\,like companion of WN8 type to the compact object in
Cygnus\,X-3, a later type than suggested by earlier works
\citep{ker96,fen99,han00}.
  
\subsection{Mass loss rate evaluation}
As \citet{ogl01}, we evaluated the mass loss rate of this free-free emitting wind, 
following the \citet{wri75} formula (8) giving the emitted flux density (in Jansky) 
by a stellar wind assumed to be spherical, homogeneous and at a constant velocity :
        \begin{displaymath}
	\centering 
        \mathrm{S}_\nu \, =\, 23.2 \, 
          \left(\frac{\dot{\mathrm{M}}}{\mu \mathrm{v}_\infty}\right)^{4/3} 
          \ \frac{\nu^{2/3}}{ \mathrm{D}_ \mathrm{kpc}^2} 
          \  \gamma^{2/3}\,\mathrm{g}^{2/3}\,\mathrm{Z}^{4/3} \quad \mathrm{Jy}
        \end{displaymath} 
        where D is the distance to the source in kpc. It gives : 
        \begin{displaymath} 
	\centering 
        \dot{\mathrm{M}}\, =\, 5.35.10^{-7} \, \mathrm{S}_\nu^{3/4}\,
                (\nu_ \mathrm{GHz}\, \gamma\, \mathrm{Z^2})^{-1/2} \,
                \mu\,\mathrm{v}_\infty \quad \mathrm{M_\odot .yr^{-1}} 
        \end{displaymath}
 (where $\nu$ is in GHz).
 With an assumed distance D\,=\,10\,kpc, a Gaunt factor g\,=\,1, a flux density 
deduced from the continuum fitting (\citealt{lut96} law, Fig.~3) of S$_\nu$\,=\,63 mJy at
 6.75\,$\mu$m (4.44$\times 10^{4}$\,GHz), 
and for a WN-type wind (where the mean atomic weight per nucleon $\mu$\,=\,1.5, 
the number of free electrons per nucleon $\gamma_e$\,=\,1 and the mean ionic charge 
Z\,=\,1), and with a velocity of v$_\infty$\,=\, 1\,500\,km.s$^{-1}$ (van Kerkwijk 1996),
 one obtains \.M\,=\,1.2$\times$$10^{-4}$\msun.yr$^{-1}$. This is in agreement with 
the mass transfer rate estimated by \citet{ker96} and (within a 
factor of 2) by \citet{ogl01}.
This result is in good agreement with the recent revised WN mass-loss rate estimates, 
which have been lowered by a factor of 2 or 3 due to clumping in the wind \citep{morr99}. 
Note that if we assume a flattened disc-like wind as reported by \citet{fen99}, 
who detected double peaked emission lines, the mass loss rate decreases but remains 
within less than a factor of 2 of that obtained in the spherical case, in all but 
extreme cases when the ratio of structural length scales exceeds about 10, 
as shown by \citet{sch82}. 
It is noticeable that \citet{chu92}, considering the radio flux from the 
southern, stellar wind component of WR\,147, derived a mass-loss rate of 
\.M\,=\,4.2$\times$$10^{-5}$\msun.yr$^{-1}$ and \citet{wil97} found a 
non spherically symetric stellar wind with a mass-loss rate of 
\.M\,=\,4.6$\times$$10^{-5}$\msun.yr$^{-1}$.
     
\subsection{Orbital modulation}
Since the length of this spectrophotometric measurement was comparable
to the 4.8\,h modulation period seen in the K-band at the level of 5$\%$
\citep{fen95}, we attempted to detect this modulation in our
data set. The points in Fig.\,6 refer to the average for the
whole (2-12\,$\mu$m) spectrum. Although, as shown in Fig.\,6, 
the measurement uncertainty of the orbitally
phase-resolved spectra was relatively high, the data clearly exclude
periodic variations of amplitude higher than 15$\%$.
\begin{figure}[!htp]
\vspace*{0.2cm} 
\centering
\includegraphics[width=5.5cm]{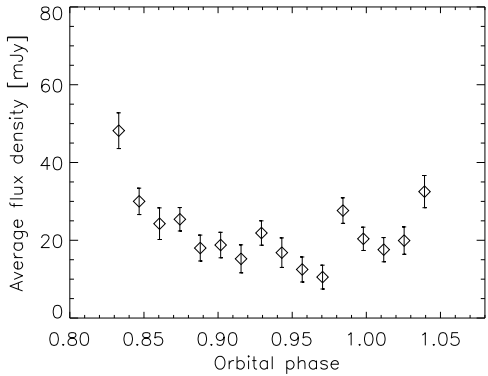}
\vspace*{1.0cm} 
\caption{Flux density in the 2.4-12 $\mu$m range versus Cygnus X-3 
orbital phase }
\label{Fig6} 
\end{figure}

\subsection{Radio and X-ray fluxes}
Figure 7 shows the mean flux density of Cygnus\,X-3 on 
MJD 50180.3 from radio to hard X-rays. The quiescent state observed 
in the mid-infrared range with ISO
was also seen during monitoring radio observations of Cygnus\,X-3 with the 
Ryle telescope (Mullard Radio Observatory, Cambridge) shortly 
before and after the ISO observations, the mean flux density at 15\,GHz being 
about 120\,mJy \citep{poo99}, and with the Green Bank Interferometer (GBI) 
monitoring program \citep{mcc99} during a quiescent period
 before the ISO observations, 
the mean flux densities being 80\,$\pm$\,30\,mJy at 2.25\,GHz and 
125\,$\pm$\,60\,mJy at 8.3\,GHz and the spectral index $\alpha$\,=\,0.3\,$\pm$\,0.1\,. 
These flux densities are at least one order of 
magnitude higher than that observed during the quench periods of very low 
radio emission preceeding the major flares of Cygnus X-3 \citep{mcc99}. In fact, this quiescent 
state was still present in 1996 May, June and July \citep{fen99}.

\begin{figure}[!htp] 
\centering
\includegraphics[width=8.7cm]{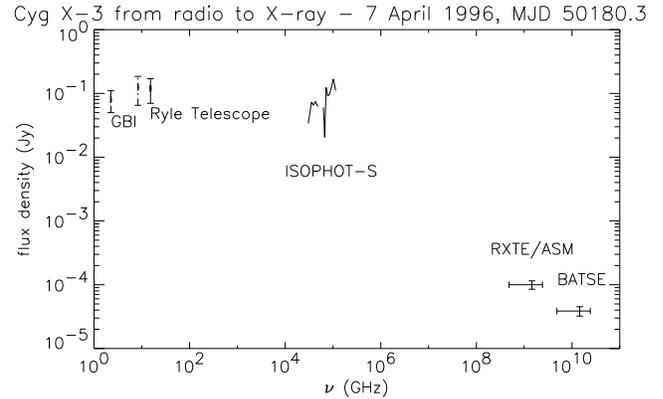}
\caption{Quasi-simultaneous observations of Cygnus\,X-3 in the radio, infrared,
 soft and hard X-rays on April\,7, 1996, averaged over the orbital phase; the 
ISOPHOT-S spectrum is dereddened with the \citet{lut96} law and rebinned to the resolution 
of 0.3\,$\mu$m;
 note that the GBI was dormant after 1996 April 1 till November 1996, and 
that the given flux densities (in dash lines) are mean values during the quiescent period 
March 17 to April 1, 1996  }
\label{Fig7} 
\end{figure}

In the X-ray range, at the same epoch, the $\it{Rossi}$ XTE/All Sky Monitor count 
rate was $\sim$\,7.5\,count.s$^{-1}$ corresponding to a
mean flux of $\sim$\,1 mJy from 2 to 12\,keV (see XTE archive and \citealt{lev96}), and 
the BATSE instrument on board the Compton Gamma Ray Observatory observed a mean photon flux 
of 0.039\,count.s$^{-1}$ corresponding to a flux density of 0.04\,mJy in the 
20\,-\,100\,keV range. Thus the mid-infrared 
continuum spectrum whose shape is explained by thermal free-free emission in an 
expanding wind has a different origin than the non-thermal radio emission
 and the hard X-ray emission which are closely coupled \citep{mio01,cho02}.

\section{Conclusions \label{Sconclusion}}
We have shown that the mid-infrared continuum (between 2.4\,-\,12\,$\mu$m) 
of Cygnus\,X-3 in quiescence 
 can be explained by the free-free emission of an 
expanding wind in the intermediate case between optically thick and optically 
thin regimes. The low quiescent luminosity of the object in the mid-infrared
allows only detection of an upper limit of 15 percent on the possible 4.8\,h
orbital modulation.   
A line at 4.3 $\mu$m is detected at a confidence level of more than 4.3\,$\sigma$, 
and is interpreted as the expected HeI (3p-3s) emission line. The close match between 
the mid-infrared brightness and spectral energy distribution of Cygnus\,X-3 in quiescence, 
the HeI emission line, the high mass loss rate in the wind and that of 
the colliding-wind Wolf-Rayet system WR\,147, is consistent with a Wolf-Rayet\,like 
companion of WN8 type to the compact object in Cygnus\,X-3, a later type than 
suggested by previous works \citep{ker96, fen99, han00}.

\begin{acknowledgements}
We warmly thank the ISO project and the ISOCAM and ISOPHOT Teams in
Villafranca, Saclay and Heidelberg. We express our gratitude to R. Ogley and 
to R. Fender for helpful comments, to G. Pooley for giving us 
the Ryle telescope data and to J.L. Starck for very useful discussions on data analysis. 
We thank the referee J. Mart\'i for helpful comments on the manuscript. 
This research has made use of data from the Green Bank Interferometer, 
a facility of the National Science Foundation operated by the NRAO in support of
 NASA High Energy Astrophysics programs, of data which were generated by the CGRO BATSE 
Instrument Team at the Marshall Space Flight Center (MSFC) using the Earth occultation 
technique, and of quick-look results provided by the ASM/RXTE team. P.A. acknowledges 
the support of a Hungarian science grant. 
\end{acknowledgements} 

%\begin{thebibliography}{}
\def\aj{AJ\ }
\def\aea{A\&A\ }
\def\aap{A\&A\ }
\def\aaps{A\&AS\ }
\def\apj{ApJ\ }
\def\apjl{ApJL\ }+
\def\apjsup{ApJS\ }
\def\mnras{MNRAS\ }
\def\nat{Nat.\ }
\def\natps{Nat.Phys.Sci.\ }
\def\sci{Sci\ }
\def\physrep{Phys. Rep.\ }
\def\apss{Ap\&SS\ }
\bibliographystyle{aa}
\bibliography{biblio}

\begin{thebibliography}{49}
\expandafter\ifx\csname natexlab\endcsname\relax\def\natexlab#1{#1}\fi

\bibitem[{{Abergel} {et~al.}(1996){Abergel}, {Bernard}, {Boulanger},
  {Cesarsky}, {Desert}, {Falgarone}, {Lagache}, {Perault}, {Puget}, {Reach},
  {Nordh}, {Olofsson}, {Huldtgren}, {Kaas}, {Andre}, {Bontemps}, {Burgdorf},
  {Copet}, {Davies}, {Montmerle}, {Persi}, \& {Sibille}}]{abe96}
{Abergel}, A., {Bernard}, J.~P., {Boulanger}, F., {et~al.} 1996, \aap, 315,
  L329

\bibitem[{{Acosta-Pulido} \& {{\' A}brah{\' a}m}(2001)}]{aco01}
{Acosta-Pulido}, J.~A. \& {{\' A}brah{\' a}m}, P. 2001, in The Calibration
  Legacy of the ISO Mission

\bibitem[{{Becklin} {et~al.}(1973){Becklin}, {Neugebauer}, {Hawkins}, {Mason},
  {Sanford}, {Mathews}, \& {Wynn-Williams}}]{bec73}
{Becklin}, E.~E., {Neugebauer}, G., {Hawkins}, F.~J., {et~al.} 1973, \nat, 245,
  302

\bibitem[{{Bonnet-Bidaud} \& {Chardin}(1988)}]{bon88}
{Bonnet-Bidaud}, J.~M. \& {Chardin}, G. 1988, \physrep, 170, 326

\bibitem[{{Cherepashchuk} \& {Moffat}(1994)}]{che94}
{Cherepashchuk}, A.~M. \& {Moffat}, A.~F.~J. 1994, \apjl, 424, L53

\bibitem[{{Choudhury} {et~al.}(2002){Choudhury}, {Rao}, {Vadawale},
  {Ishwara-Chandra}, \& {Jain}}]{cho02}
{Choudhury}, M., {Rao}, A.~R., {Vadawale}, S.~V., {Ishwara-Chandra}, C.~H., \&
  {Jain}, A.~K. 2002, \aap, 383, L35

\bibitem[{{Churchwell} {et~al.}(1992){Churchwell}, {Bieging}, {van der Hucht},
  {Williams}, {Spoelstra}, \& {Abbott}}]{chu92}
{Churchwell}, E., {Bieging}, J.~H., {van der Hucht}, K.~A., {et~al.} 1992,
  \apj, 393, 329

\bibitem[{{Crowther} {et~al.}(1995{\natexlab{a}}){Crowther}, {Hillier}, \&
  {Smith}}]{cro95II}
{Crowther}, P.~A., {Hillier}, D.~J., \& {Smith}, L.~J. 1995{\natexlab{a}},
  \aap, 293, 403

\bibitem[{{Crowther} {et~al.}(1995{\natexlab{b}}){Crowther}, {Smith}, \&
  {Hillier}}]{cro95IV}
{Crowther}, P.~A., {Smith}, L.~J., \& {Hillier}, D.~J. 1995{\natexlab{b}},
  \aap, 302, 457

\bibitem[{{Draine}(1989)}]{dra89}
{Draine}, B.~T. 1989, in Infrared Spectroscopy in Astronomy, 93--98

\bibitem[{{Ergma} \& {van den Heuvel}(1998)}]{erg98}
{Ergma}, E. \& {van den Heuvel}, E.~P.~J. 1998, \aap, 331, L29

\bibitem[{{Fender} {et~al.}(1995){Fender}, {Bell Burnell}, {Garrington},
  {Spencer}, \& {Pooley}}]{fen95}
{Fender}, R.~P., {Bell Burnell}, S.~J., {Garrington}, S.~T., {Spencer}, R.~E.,
  \& {Pooley}, G.~G. 1995, \mnras, 274, 633

\bibitem[{{Fender} {et~al.}(1996){Fender}, {Bell Burnell}, {Williams}, \&
  {Webster}}]{fen96}
{Fender}, R.~P., {Bell Burnell}, S.~J., {Williams}, P.~M., \& {Webster}, A.~S.
  1996, \mnras, 283, 798

\bibitem[{{Fender} {et~al.}(1999){Fender}, {Hanson}, \& {Pooley}}]{fen99}
{Fender}, R.~P., {Hanson}, M.~M., \& {Pooley}, G.~G. 1999, \mnras, 308, 473

\bibitem[{{Gabriel} {et~al.}(1997){Gabriel}, {Acosta-Pulido}, {Heinrichsen},
  {Morris}, \& {Tai}}]{gab97}
{Gabriel}, C., {Acosta-Pulido}, J., {Heinrichsen}, I., {Morris}, H., \& {Tai},
  W.-M. 1997, in ASP Conf. Ser. 125: Astronomical Data Analysis Software and
  Systems VI, Vol.~6, 108--

\bibitem[{{Giacconi} {et~al.}(1967){Giacconi}, {Gorenstein}, {Gursky}, \&
  {Waters}}]{gia67}
{Giacconi}, R., {Gorenstein}, P., {Gursky}, H., \& {Waters}, J.~R. 1967, \apjl,
  148, L119

\bibitem[{{Haas} {et~al.}(2000){Haas}, {M{\" u}ller}, {Chini}, {Meisenheimer},
  {Klaas}, {Lemke}, {Kreysa}, \& {Camenzind}}]{haa00}
{Haas}, M., {M{\" u}ller}, S.~A.~H., {Chini}, R., {et~al.} 2000, \aap, 354, 453

\bibitem[{{Hanson} {et~al.}(2000){Hanson}, {Still}, \& {Fender}}]{han00}
{Hanson}, M.~M., {Still}, M.~D., \& {Fender}, R.~P. 2000, \apj, 541, 308

\bibitem[{{Kessler} {et~al.}(1996){Kessler}, {Steinz}, {Anderegg}, {Clavel},
  {Drechsel}, {Estaria}, {Faelker}, {Riedinger}, {Robson}, {Taylor}, \&
  {Ximenez de Ferran}}]{kes96}
{Kessler}, M.~F., {Steinz}, J.~A., {Anderegg}, M.~E., {et~al.} 1996, \aap, 315,
  L27

\bibitem[{{Kitamoto} {et~al.}(1992){Kitamoto}, {Mizobuchi}, {Yamashita}, \&
  {Nakamura}}]{kit92}
{Kitamoto}, S., {Mizobuchi}, S., {Yamashita}, K., \& {Nakamura}, H. 1992, \apj,
  384, 263

\bibitem[{{Klaas} {et~al.}(1994){Klaas}, {Kr\"uger}, {Henrichsen}, \&
  {Laureijs}}]{kla94}
{Klaas}, U., {Kr\"uger}, H., {Henrichsen}, I., \& {Laureijs}, R. 1994, ISOPHOT
  Observers Manual, Version 3.1, ISO Science Operations Team, ESA/ESTEC,
  Noordwijk, The Netherlands, http://www.iso.vilspa.esa.es/manuals/

\bibitem[{{Koch-Miramond} {et~al.}(2001){Koch-Miramond}, {Bonnet-Bidaud},
  {Abraham}, \& {Claret}}]{koc01}
{Koch-Miramond}, L., {Bonnet-Bidaud}, J., {Abraham}, P., \& {Claret}, A. 2001,
  in Black Holes in Binaries and Galactic Nuclei, eds L. Kaper, E.P.J. van den
  Heuvel \& P.A. Woudt, ESO Astrophysics Symposia, 137

\bibitem[{{Levine} {et~al.}(1996){Levine}, {Bradt}, {Cui}, {Jernigan},
  {Morgan}, {Remillard}, {Shirey}, \& {Smith}}]{lev96}
{Levine}, A.~M., {Bradt}, H., {Cui}, W., {et~al.} 1996, \apjl, 469, L33

\bibitem[{{Lutz} {et~al.}(1996){Lutz}, {Feuchtgruber}, {Genzel}, {Kunze},
  {Rigopoulou}, {Spoon}, {Wright}, {Egami}, {Katterloher}, {Sturm},
  {Wieprecht}, {Sternberg}, {Moorwood}, \& {de Graauw}}]{lut96}
{Lutz}, D., {Feuchtgruber}, H., {Genzel}, R., {et~al.} 1996, \aap, 315, L269

\bibitem[{{Mart{\' i}} {et~al.}(2000){Mart{\' i}}, {Paredes}, \&
  {Peracaula}}]{mar00}
{Mart{\' i}}, J., {Paredes}, J.~M., \& {Peracaula}, M. 2000, \apj, 545, 939

\bibitem[{{Mart{\' i}} {et~al.}(2001){Mart{\' i}}, {Paredes}, \&
  {Peracaula}}]{mar01}
---. 2001, \aap, 375, 476

\bibitem[{{Mason} {et~al.}(1986){Mason}, {Cordova}, \& {White}}]{mas86}
{Mason}, K.~O., {Cordova}, F.~A., \& {White}, N.~E. 1986, \apj, 309, 700

\bibitem[{{McCollough} {et~al.}(1999){McCollough}, {Robinson}, {Zhang},
  {Harmon}, {Hjellming}, {Waltman}, {Foster}, {Ghigo}, {Briggs}, {Pendleton},
  \& {Johnston}}]{mcc99}
{McCollough}, M.~L., {Robinson}, C.~R., {Zhang}, S.~N., {et~al.} 1999, \apj,
  517, 951

\bibitem[{{Mioduszewski} {et~al.}(1998){Mioduszewski}, {Hjellming}, {Rupen},
  {Waltman}, {Pooley}, {Ghigo}, \& {Fender}}]{mio98}
{Mioduszewski}, A.~J., {Hjellming}, R.~M., {Rupen}, M.~P., {et~al.} 1998, in
  ASP Conf. Ser. 144: IAU Colloq. 164: Radio Emission from Galactic and
  Extragalactic Compact Sources, 351

\bibitem[{{Mioduszewski} {et~al.}(2001){Mioduszewski}, {Rupen}, {Hjellming},
  {Pooley}, \& {Waltman}}]{mio01}
{Mioduszewski}, A.~J., {Rupen}, M.~P., {Hjellming}, R.~M., {Pooley}, G.~G., \&
  {Waltman}, E.~B. 2001, \apj, 553, 766

\bibitem[{{Mitra}(1996)}]{mit96}
{Mitra}, A. 1996, \mnras, 280, 953

\bibitem[{{Morel} {et~al.}(1999){Morel}, {Marchenko}, {Eenens}, {Moffat},
  {Koenigsberger}, {Antokhin}, {Eversberg}, {Tovmassian}, {Hill}, {Cardona}, \&
  {St-Louis}}]{more99}
{Morel}, T., {Marchenko}, S.~V., {Eenens}, P.~R.~J., {et~al.} 1999, \apj, 518,
  428

\bibitem[{{Morris} {et~al.}(2000){Morris}, {van der Hucht}, {Crowther},
  {Hillier}, {Dessart}, {Williams}, \& {Willis}}]{morr00}
{Morris}, P.~W., {van der Hucht}, K.~A., {Crowther}, P.~A., {et~al.} 2000,
  \aap, 353, 624

\bibitem[{{Morris} {et~al.}(1999){Morris}, {van der Hucht}, {Willis},
  {Dessart}, {Crowther}, \& {Williams}}]{morr99}
{Morris}, P.~W., {van der Hucht}, K.~A., {Willis}, A.~J., {et~al.} 1999, in IAU
  Symp. 193: Wolf-Rayet Phenomena in Massive Stars and Starburst Galaxies, Vol.
  193, 77--

\bibitem[{{Ogley} {et~al.}(2001){Ogley}, {Bell Burnell}, \& {Fender}}]{ogl01}
{Ogley}, R.~N., {Bell Burnell}, S.~J., \& {Fender}, R.~P. 2001, \mnras, 322,
  177

\bibitem[{Parsignault(1972)}]{par72}
Parsignault, D.~R., e.~a. 1972, \natps

\bibitem[{{Pittard} {et~al.}(2002){Pittard}, {Stevens}, {Williams}, {Pollock},
  {Skinner}, {Corcoran}, \& {Moffat}}]{pit02}
{Pittard}, J.~M., {Stevens}, I.~R., {Williams}, P.~M., {et~al.} 2002, \aap,
  388, 335

\bibitem[{Pooley(1999)}]{poo99}
Pooley, G.~G. 1999, private communication

\bibitem[{{Schmid-Burgk}(1982)}]{sch82}
{Schmid-Burgk}, J. 1982, \aap, 108, 169

\bibitem[{{Skinner} {et~al.}(1999){Skinner}, {Itoh}, {Nagase}, \&
  {Zhekov}}]{ski99}
{Skinner}, S.~L., {Itoh}, M., {Nagase}, F., \& {Zhekov}, S.~A. 1999, \apj, 524,
  394

\bibitem[{{Starck} {et~al.}(1999){Starck}, {Abergel}, {Aussel}, {Sauvage},
  {Gastaud}, {Claret}, {Desert}, {Delattre}, \& {Pantin}}]{sta99}
{Starck}, J.~L., {Abergel}, A., {Aussel}, H., {et~al.} 1999, \aaps, 134, 135

\bibitem[{{Stevens} \& {Howarth}(1999)}]{ste99}
{Stevens}, I.~R. \& {Howarth}, I.~D. 1999, \mnras, 302, 549

\bibitem[{{van den Heuvel} \& {de Loore}(1973)}]{heu73}
{van den Heuvel}, E.~P.~J. \& {de Loore}, C. 1973, \aap, 25, 387

\bibitem[{{van der Hucht}(2001)}]{huc01}
{van der Hucht}, K.~A. 2001, New Astronomy Review, 45, 135

\bibitem[{{van Kerkwijk} {et~al.}(1992){van Kerkwijk}, {Charles}, {Geballe},
  {King}, {Miley}, {Molnar}, {van den Heuvel}, {van der Klis}, \& {van
  Paradijs}}]{ker92}
{van Kerkwijk}, M.~H., {Charles}, P.~A., {Geballe}, T.~R., {et~al.} 1992, \nat,
  355, 703

\bibitem[{{van Kerkwijk} {et~al.}(1996){van Kerkwijk}, {Geballe}, {King}, {van
  der Klis}, \& {van Paradijs}}]{ker96}
{van Kerkwijk}, M.~H., {Geballe}, T.~R., {King}, D.~L., {van der Klis}, M., \&
  {van Paradijs}, J. 1996, \aap, 314, 521

\bibitem[{{Vanbeveren} {et~al.}(1998){Vanbeveren}, {de Donder}, {van Bever},
  {van Rensbergen}, \& {de Loore}}]{van98}
{Vanbeveren}, D., {de Donder}, E., {van Bever}, J., {van Rensbergen}, W., \&
  {de Loore}, C. 1998, New Astronomy, 3, 443

\bibitem[{{Williams} {et~al.}(1997){Williams}, {Dougherty}, {Davis}, {van der
  Hucht}, {Bode}, \& {Setia Gunawan}}]{wil97}
{Williams}, P.~M., {Dougherty}, S.~M., {Davis}, R.~J., {et~al.} 1997, \mnras,
  289, 10

\bibitem[{{Wright} \& {Barlow}(1975)}]{wri75}
{Wright}, A.~E. \& {Barlow}, M.~J. 1975, \mnras, 170, 41

\end{thebibliography}

%\end{thebibliography}

\end{document}